\documentclass[prb,twocolumn,superscriptaddress,showpacs,preprintnumbers,amsmath,amssymb]{revtex4}
\usepackage{graphicx}
\usepackage{dcolumn}
\usepackage{bm}
\usepackage{mathrsfs}
\usepackage{subfigure}
\usepackage{natbib}

\begin{document}

\title{Stabilized silicene within bilayer graphene: A proposal based on molecular dynamics
and density-functional tight-binding calculations}

\author{G. R. Berdiyorov}
\affiliation{Department of Physics, King Fahd University of Petroleum and Minerals, 31261 Dhahran, Saudi Arabia}
\affiliation{Departement Fysica, Universiteit Antwerpen, Groenenborgerlaan 171, B-2020 Antwerpen, Belgium}
\author{M. Neek-Amal}
\affiliation{Departement Fysica, Universiteit Antwerpen, Groenenborgerlaan 171, B-2020 Antwerpen, Belgium}
\affiliation{Department of Physics, Shahid Rajaee Teacher Training University, Lavizan, Tehran 16785-136, Iran}
\author{F. M. Peeters}
\email{francois.peeters@uantwerpen.be} \affiliation{Departement Fysica, Universiteit Antwerpen, Groenenborgerlaan 171,
B-2020 Antwerpen, Belgium}
\author{Adri C. T. van Duin}
\affiliation{Department of Mechanical and Nuclear Engineering, the Pennsylvania State University, University Park, PA 16802, USA}

\date{\today}

\begin{abstract}

Free standing silicene is predicted to display comparable electronic properties as graphene. However, the yet synthesized silicene-like structures have been only realized on different substrates which turned out to exhibit versatile crystallographic structures that are very different from the theoretically predicted buckled phase of freestanding silicene. This calls for a different approach where silicene is stabilized using very weakly interacting surfaces. We propose here a novel route by using graphene bilayer as a scaffold. The confinement between the flat graphene layers results in a planar clustering of Si atoms with small buckling, which is energetically unfavorable in vacuum. Buckled hexagonal arrangement of Si atoms similar to free-standing silicene is observed for large clusters, which, in contrast to Si atoms on metallic surfaces, is only very weakly van der Waals coupled to the graphene layers. These clusters are found to be stable well above room temperature. Our findings, which are supported by density functional tight-binding calculations, show that intercalating bilayer graphene with Si is a favorable route to realize silicene.

\end{abstract}

\maketitle

\section{Introduction}

The mechanical exfoliation of graphene \cite{Novoselov2005} from graphite has resulted in an enormous interest in this two-dimensional monolayer of hexagonal ordered carbon atoms which is due to its unique electronic, optical and mechanical properties \cite{Castro2009} that is expected to lead to potential applications in different areas of electronics, opto-electronics, etc. A material of even more current technological importance is silicon. It has been a longstanding debate, dating back to the pioneering work of Yin and Cohen,\cite{Yin1984} whether or not graphitic Si is stable. Using pseudopotential local-density-functional theory they doubted that the formation of graphitic Si is possible because its energy is 0.71 eV/atom higher than the diamond phase and that a large negative pressure of -69 kbar is needed to stabilize it. Even before the exfoliation of graphene in 2004 several theoretical works \cite{Takeda1994,Fagan2000} have predicted that free-standing single-layer silicon, called silicene,\cite{Zhang2005,Guzman2007,Cahangirov2009,SahinPRB,Houssa2010} is stable. Graphene's sp$^2$ hybridization leads to a flat layer, which in silicene is unfavorable with respect to a buckled Si(111) honeycomb structure with sp$^3$ hybridization. The electronic structure has been shown to be similar to graphene that is characterized by a zero gap and a Dirac cone low energy spectrum. A unique feature of silicene is its large spin-orbit interaction \cite{LiuPRB2011} that is predicted to result in quantum spin Hall effect,\cite{LiuPRL2011} electrically tunable band gap \cite{Drummond2012} and the emergence of a valley-polarized metal phase.\cite{Ezawa2012}

Recently there has been very active experimental research on the synthesis of silicene. A versatility of different silicene-like crystallographic structures were obtained on different substrates, e.g. Ag,\cite{Ag} ZrB$_2$,\cite{Zr} Ir,\cite{Ir})%
depending on the growth conditions and the particular arrangement of the substrate atoms. These phases were found to be different from the theoretical predicted buckled configuration of freestanding silicene which are a consequence of the interaction with the substrate and the induced epitaxial strain. Furthermore, there is yet no definite proof that these silicene like structures exhibit a zero gap with a Dirac cone electronic spectrum.\cite{Lin2013,Guo2013,Wang2013,Cahangirov2013} Up to now no freestanding silicene has been fabricated and there is serious doubt that such a structure is even stable in nature. Here we propose that alternatively, one can insert silicene between two substrates that interact very weakly with the Si atoms, in order to stabilize it. We show that once intercalated into the bilayer graphene (either through domain boundaries or vacancy defects in graphene layers \cite{Kaloni}), the silicon atoms can be stabilized to planar (with small buckling) silicon clusters during thermal annealing, which can serve as building blocks for a silicene sheet. It has been shown in recent density functional theory (DFT) calculations that silicene layer strengthens the interlayer binding between the graphene sheets as compared to that in graphite without altering the other properties of graphene like the Dirac fermion-like electronic structure.\cite{Zhang2010}

In this work we conduct a systematic study of the structural properties and the thermal stability of Si atoms intercalating bilayer graphene using reactive molecular dynamics (MD) simulations, which are supported by density functional tight-binding theory (DFTB).\cite{Porezag, Elstner,Frauenheim} We found that energetically unfavorable planar silicon clusters in free space can be stabilized to slightly buckled honeycomb structures by the weak confinement due to the induced straining in the graphene layers. These quasi-two-dimensional (2D) Si clusters are found to be different from those of silicene over an Ag substrate \cite{Gao_nature} and are stable beyond room temperature. At higher temperature they transit to three-dimensional (3D) diamond-like structures with predominant sp$^3$ hybridization. Since such structures can naturally arise during epitaxial growth of few-layer graphene on bulk silicon carbide (SiC) by thermal decomposition~\cite{Berger2006}, our findings can be useful in the understanding of the mechanisms for synthesis of multilayer graphene on SiC.\cite{Singh2011,Xia2012,Wang2012} The results may also initiate further research on graphene-silicene superlattice structures with promising structural and electronic properties. Recently, it has been proposed \cite{Jose2011} that analogous to graphene \cite{Sofo2007} hydrogenation of silicene clusters maybe a promising route for hydrogen storage.

\section{Computational method}

To study the structural (thermal) properties of silicene, MD simulations were performed using the reactive force-field ReaxFF, which, in contrast to classical force-fields, is a general bond-order dependent potential that accounts for bond breaking and bond formation during chemical reactions.\cite{vanDuin} The
system connectivity is recalculated at every iteration step and non-bounded interactions (van der Waals and Coulomb) are calculated between all atom pairs, irrespective of their connectivity.\cite{vanDuin,Newsome,PRB2013} Since ReaxFF parameters are derived from quantum chemical calculations, it gives
energies, transition states, reaction pathways and reactivity trends in agreement with quantum mechanical calculations and experiments.\cite{vanDuin} Numerical simulations are carried out using the LAMMPS code~\cite{Kresse} which includes the ReaxFF method.\cite{Zybin} In order to have an independent test of our used model we performed extra calculations using DFTB/MD ~\cite{Porezag,Elstner, Frauenheim}.


\begin{figure}[t]
\includegraphics[width=\linewidth]{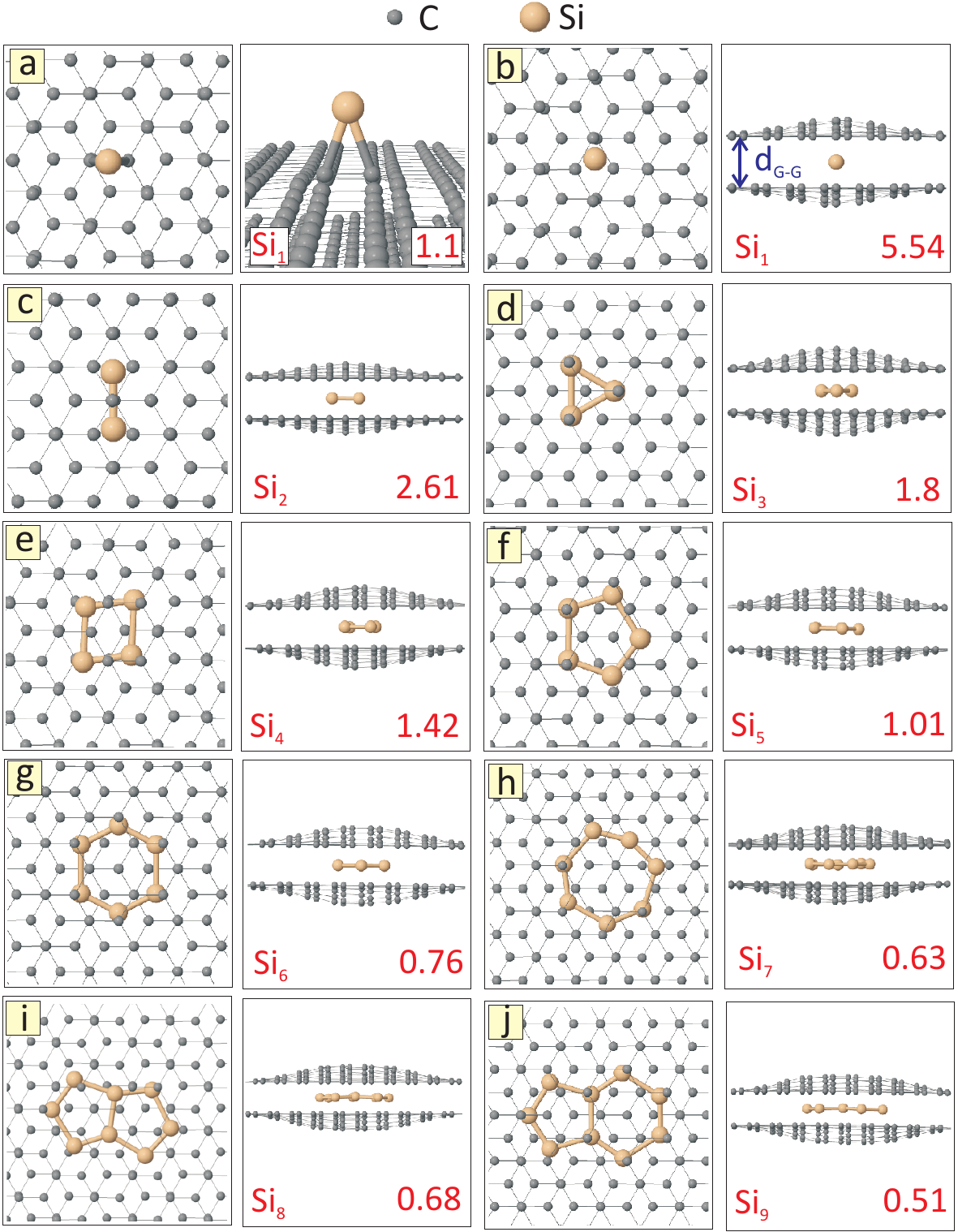}
\caption{\label{fig1} (color online) (a) A single Si atom on top of bilayer graphene. (b-j) Ground state configurations [top (side) view on the left (right)] of Si$_N$ clusters intercalating bilayer graphene. The numbers show the formation energies in eV per Si atom.}
\end{figure}

\section{Structural properties}

We present now a systematic study of the structural properties of Si atoms inserted between the graphene layers. Such intercalation can occur e.g. through domain boundaries or defect areas in the graphene layer.\cite{Kaloni} Here we report only the energetically most favorable configurations out of the many possible metastable configurations (including 3D Si clusters) which we investigated. As a representative example, we consider AB stacked bilayer graphene (with 960 carbon atoms in each layer, corresponding to a computational unit cell of 5.31$\times$5.01 nm$^2$ in the $x-y$-plane) with periodic boundary conditions along the graphene basal planes in order to avoid edge effects. The formation energies of Si clusters are calculated as:\cite{Gao_nature} $E_f=(E_t-N\times E_{Si}-E_g)/N$, where $E_t$ is the total energy of the system, $N$ is the number of atoms in the cluster, $E_{Si}=4.63$ eV is the cohesive energy of Si, and $E_g$ is the energy of bilayer graphene.

We start by considering a single silicon atom adsorbed on top of bilayer graphene, the equilibrium structure of which is shown in Fig. \ref{fig1}(a). The Si atom is adsorbed at bridge site forming a covalent bond (above the middle of the carbon-carbon bonds) with bond distance $d_{Si-C}=2.08 \AA$ and formation energy of 1.1 eV. Such bridge site attachment was recently predicted using first principles calculations with binding energy of 1.17 eV and Si-C distance in the range $d_{Si-C}=2.04-2.11 \AA$.\cite{Akturk} Si attachment induces a small local change in the underlaying planar graphene [see right panel of Fig. \ref{fig1}(a)] where the carbon-carbon distance increases to $1.47\AA$, which is very close to the DFT prediction ($1.45\AA$).\cite{Akturk} This result is also very different from the assumed hollow side positioning of the Si atoms between graphene layers proposed in Ref. \cite{Zhang2010}.

Figure \ref{fig1}(b) shows the ground state configuration of the system when the Si atom is inserted between the layers. This Si atom is located at equal distance from both layers and results in considerable local expansion and buckling of the layers. The interlayer distance of bilayer graphene without intercalating bilayer graphene is 3.29$\AA$. The local deformation of the graphene layers (in this particular case about 50\%) explains the considerable enhancement of the formation energy as compared to the case when the Si atom is adsorbed on top of bilayer graphene. $E_f$ decreases more than twice by the formation of a silicon dimer [Fig.~\ref{fig1}(c)], which is also located in the middle of the interplanar spacing and weakly van der Waals bonded with the graphene layers. A triangular cluster is found in the ground state for $N$=3 as in the case of Si$_3$ on a metallic surface.\cite{Gao_nature}. All three Si atoms are located at the center of the hexagonal ring of the lower graphene layer and right below the carbon atom of the upper layer, i.e., so called the ``H-T'' site [Fig. \ref{fig1}(d)].\cite{Xian} Such energy-minimum locations between the layers are observed for larger Si clusters [see Figs. \ref{fig1}(e,f)]. Notice that line structures observed for free standing carbon clusters C$_N$ with $N\le 5$ (see e.g., Ref. \cite{Singh2013}) are found to be metastable in case of Si clusters. Figure \ref{fig1}(g) shows the equilibrium state of Si$_6$, which is the building block for silicene. Adding one Si atom to it results in the Si$_7$ cluster which shows (Fig. \ref{fig1}(h)) a planar ring structure with different Si interatomic distances. The reason is that the Si atoms try to accommodate the graphene matrix. The single ring structure becomes energetically less stable with further increasing $N$, and a double ring structure is found for $N=8$ and $N=9$, as shown in Figs. \ref{fig1}(i,j). All the considered ground state structures are planar except for Si$_8$ and Si$_9$, for which we observed a slight buckling (less than 0.2 $\AA$). Note that the formation energy decreases almost monotonically by increasing the number $N$ of Si atoms in the cluster. Thus, we found that \textit{the structure of Si clusters inside bilayer graphene is totally different from free standing Si clusters} (see, e.g., Ref. \cite{Liu1998}) \textit{and for $N>$5 resembles (but are not identical) the ones observed for carbon clusters in vacuum}~\cite{Singh2013}.
\begin{figure}[b]
\includegraphics[width=\linewidth]{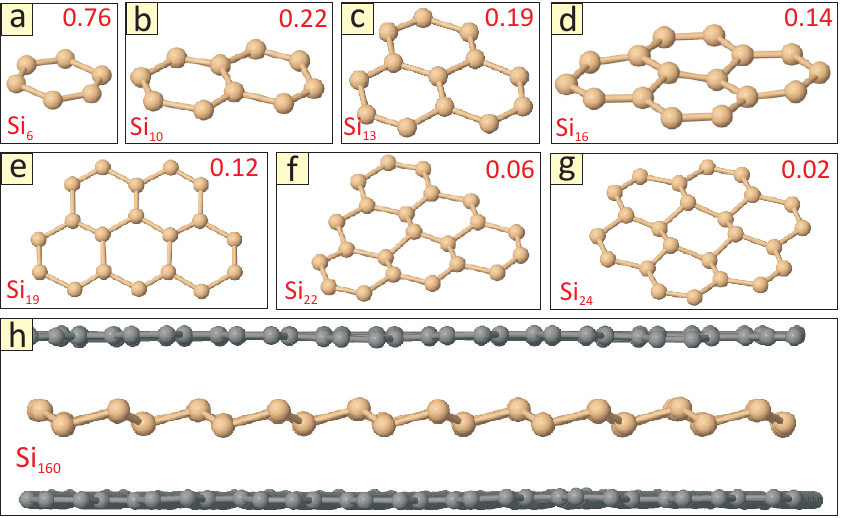}
\caption{\label{fig2} (color online) (a-g) Geometries and formation energies (eV per Si atom)
of Si clusters with six-membered rings inside a bilayer graphene (graphene layers are not shown). (h) Silicene sheet (160 Si atom computational unit cell) intercalating bilayer graphene.}
\end{figure}

\begin{figure}[b]
\includegraphics[width=\linewidth]{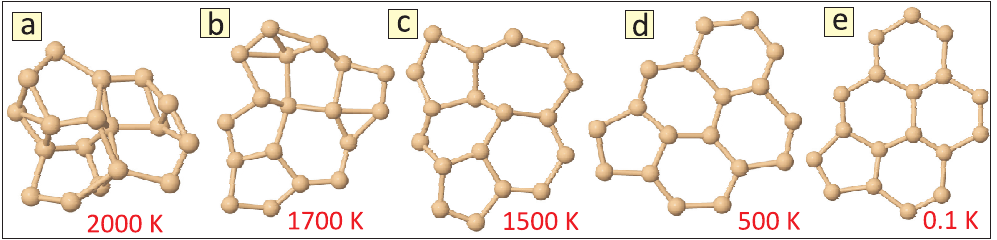}
\caption{\label{fig3} (color online) Snapshots of Si$_{18}$ cluster between bilayer graphene at different temperatures during a rapid cooling.}
\end{figure}

In what follows, we study the structural properties of Si$_N$ clusters consisting of six-membered silicon rings. Initial atomic configurations with a planar honeycomb arrangement of Si atoms are used (see, e.g. Ref. \cite{Gao_nature}). Equilibrium geometries and formation energies are shown in Figs. \ref{fig2}(a-g) for $N=$6, 10, 13, 16, 19, 22 and 24. Interestingly, all the clusters retain their hexagonal ring structure upon relaxation, whereas earlier work of Si on metallic surfaces found that some of the clusters changes its structure.\cite{Gao_nature} As in the case of smaller Si$_N$ clusters (see Fig. \ref{fig1}), the Si atoms are not equidistant and the buckling becomes more pronounced with increasing $N$. However, the buckling is not constant across the cluster: it is larger in the middle of the cluster and decreases towards the edges. As in the case of metal supported Si clusters \cite{Gao_nature}, the formation energy of Si$_N$ clusters decreases as the cluster size increases.

Figure \ref{fig2}(h) shows the equilibrium structure of a silicene sheet (with 160 Si atoms in the unit cell) intercalating bilayer graphene (360 C atoms in each layer of the computational unit cell). We started our simulations with a flat hexagonal arrangement of silicon atoms -- the state which was shown to be unstable in DFT simulations.\cite{Cahangirov2009} The system transits to a buckled structure upon energy minimization as shown in Fig. \ref{fig2}(h). In the optimized geometry the averaged inter-atomic distance is around 2.24 $\AA$, which is close to the DFT predictions for silicene in vacuum (2.25 $\AA$).\cite{Cahangirov2009} The averaged buckling parameter equals $\sigma$=0.65$\pm$0.07 $\AA$, which is comparable to the buckling of silicene on an Ag(111) surface $\sigma$=0.85~\AA.\cite{Gao_nature} The distance between the graphene layers is $d_{G-G}$=6.92 \AA, which is almost twice larger than the interlayer spacing in graphite. Such stacked structure of planar graphene and buckled silicene layers has promising applications due to the fact that the properties of both graphene and silicene remains unaltered, i.e. both silicene and graphene exhibit a Dirac cone at the $K$-point but with their respective Dirac points displaced energy.\cite{Mehdi}

When studying larger size Si clusters inside bilayer graphene (see Fig. \ref{fig2}), we have a predetermined hexagonal arrangement of Si atoms. However, in a real experiments Si atoms may intercalate between the graphene layers only at high temperatures (above 1000 K),\cite{Xia2012} where silicene is predicted to be unstable.\cite{Cahangirov2009} Thus the interesting question in this case is what happens to the Si cluster when temperature is rapidly quenched. To model this situation, we conducted the following simulations: first, we initialized the Si atoms randomly far from each other, then increased temperature gradually (20 K/ps) from 0 to 2000 K. During this process Si atoms start migrating (due to their small migration barrier~\cite{Xian}) and forms a 3D-like cluster at high temperatures, as shown in Fig. \ref{fig3}(a). After that we decreased the temperature with 20 K/ps. The cluster rearranges itself into different irregular shapes during this process [Figs. \ref{fig3}(b,c)]. At temperatures below 500 K the cluster is transformed into six-member Si rings [Fig. \ref{fig3}(d)] and for lower temperatures we observe the same planar cluster [with slight buckling, see Fig. \ref{fig3}(e)] as we reported in Fig. \ref{fig2}. Animated online video \cite{suppl_1} shows such structural transformation. Note that the formation of such Si cluster does not depend on the rate of temperature increase/decrease. Thus, we predict that \textit{planar Si clusters can be formed in real experiments} provided that Si atoms intercalate between the graphene layers.

\begin{figure}[t]
\includegraphics[width=\linewidth]{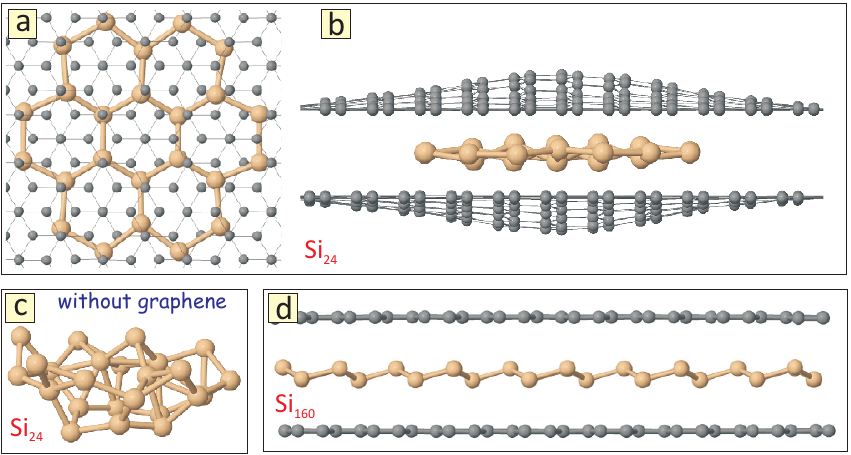}
\caption{\label{fig4} (color online) DFTB results: Top (a) and side (b) view of the equilibrium structure of Si$_{24}$ cluster intercalated inside bilayer graphene with 960 carbon atoms in each layer. (c) Optimized structure of Si$_{24}$ cluster without graphene. (d) Silicene sheet (160 Si atoms) intercalating bilayer graphene.}
\end{figure}

\section{DFTB calculations}

To support our findings about the stability of planar (with small buckling) Si clusters intercalated bilayer graphene, we conducted simulations using DFTB theory, which is approximately two orders of magnitude faster than DFT and therefore enables one to model larger systems.\cite{Porezag,Elstner,Frauenheim} As a typical example, we consider a Si$_{24}$ cluster intercalating bilayer graphene with 960 carbon atoms in each layer, i.e., the same system as in Fig. \ref{fig2}(g). Figures \ref{fig4}(a,b) show the optimized structure of the system, which shows the same buckled structure as found in our MD simulations [see Fig. \ref{fig2}(g)]. Notice that we started from a random distribution of Si atoms in a 2D plane (see supplemental online video \cite{suppl_2}). In the equilibrium state, the maximal buckling of the Si$_{24}$ cluster is $\sigma=0.719~\AA$ and the maximal deformation of the graphene bilayer is 107\%. The predictions of our MD simulations for these parameters are $\sigma=0.715~\AA$ and 103\%. Note that such planar structures are not even metastable in vacuum and transform spontaneously into severely buckled configurations upon optimization, as shown in Fig. \ref{fig4}(c), in accordance with our DFT calculations.\cite{Gao_nature} Figure \ref{fig4}(d) shows the optimized structure of silicene (160 Si atoms in the unit cell) intercalating bilayer graphene (360 carbon atoms in each layer), i.e., the same system as in Fig. \ref{fig2}(h). Graphene layers preserve their planar structure during optimization and the buckling of the silicene layer is $\sigma=0.67\pm0.05\AA$, which is close to the predictions of our MD simulations ($\sigma=0.65\pm0.07\AA$). Thus, DFTB simulations \textit{confirm independently the stability of the nearly planar honeycomb arranged Si clusters intercalated bilayer graphene}.

\begin{figure}[b]
\includegraphics[width=\linewidth]{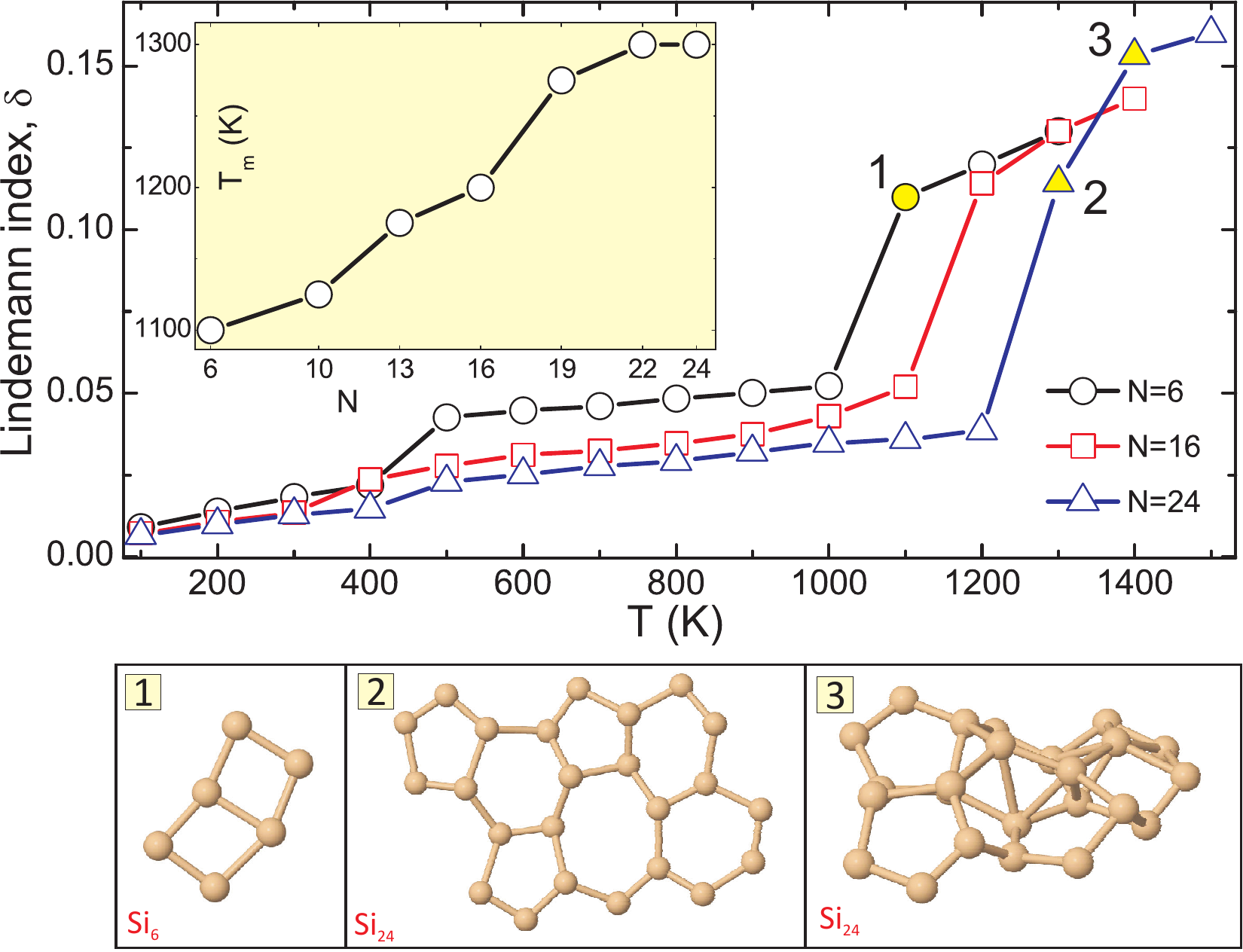}
\caption{\label{fig5} (color online) Lindemann index of Si$_N$ clusters intercalating bilayer graphene as a function of temperature for different $N$. Inset shows the melting temperature $T_m$ of quasi-2D Si$_N$ clusters constructed using the criterion $\delta=0.1$. Panels 1-3 show snapshots of the Si$_N$ clusters at temperatures indicated in the main panel indicated by filled symbols.}
\end{figure}

\section{Thermal stability}

Finally, we consider the thermal stability of the hexagon-based structures that we reported in Fig.~\ref{fig2}. Starting from the equilibrium state, we increased the temperature of the system up to 2000 K at a rate of 4 K/ps using an isothermalisobaric (NPT) ensemble with a Nose-Hoover thermostat/barostat for
temperature/pressure control. When the desired temperature is reached, constant temperature MD simulations were performed during 500 ps.\cite{details} To characterize the thermal stability of Si$_N$ clusters, we monitored the bond length fluctuations given by the Lindemann index,
\begin{eqnarray}
\delta_i=\frac{1}{N-1}\sum_{j\ne i}\sqrt{\langle r^2_{i,j}\rangle-\langle r_{i,j}\rangle^2}\Big/\langle r_{i,j}\rangle,~~~\delta=\frac{1}{N}\sum_i\delta_i,\nonumber
\end{eqnarray}
where $N$ is the number of Si atoms in the cluster, $\delta_i$ is the Lindemann index of atom $i$ and $\delta$ is the Lindemann index for the entire cluster~\cite{Singh2013}. Figure \ref{fig5} shows the Lindemann index of Si$_6$ (circles), Si$_{16}$ (squares) and Si$_{24}$ (triangles) clusters as a function of temperature. As a general trend for small clusters (see e.g. Ref. \cite{PRB2013}), $\delta$ increases linearly with temperature for low temperatures. A clear jump is observed in the $\delta(T)$ at higher temperatures, which corresponds to the formation of defects in the honeycomb structure of Si atoms (see panels 1 and 2 of Fig. \ref{fig5}). With further increasing temperature a transition of quasi-planar silicon clusters into 3D-clusters is observed (panel 3 of Fig. \ref{fig5}). Using the criterion of $\delta=0.1$, we calculated the temperature $T_m$ at which a structural transformation of the silicene clusters into 3D-like clusters takes place. The inset of Fig. \ref{fig5} shows that $T_m$ increases with increasing number of 6-member rings in the system and that $T_m$ tends to saturate around 1300 K for larger $N$. In spite of the fact that free standing planar Si clusters are not stable [see Fig. \ref{fig4}(c)], here all the quasi-2D clusters are found to be stable well above room temperature, indicating \textit{the considerable contribution of graphene layers to the stability of silicene} Note, however, that no structural transformation of the graphene matrix is observed for the considered range of temperatures ($T\le 2000$K) in our 500 ps-long simulations.

\section{Conclusions}

Using reactive molecular dynamics and DFTB simulations, we studied systematically the structural properties and thermal stability of Si atoms intercalating bilayer graphene. Due to the confinement from the graphene layers, Si atoms form planar clusters, which are energetically unfavorable for free standing Si clusters. Large Si clusters form a buckled honeycomb structure resembling the properties of free standing silicene predicted by first principles calculations.\cite{Cahangirov2009} Our simulations show that silicene intercalating graphene layers is much closer to pristine silicene than silicene on metallic surfaces because of the very small van der Waals interaction of graphene on the silicene crystal structure. Therefore, graphene layers are an almost ideal template for the formation of silicene. Silicon clusters intercalating multilayers of graphene have the potential for designing high-capacity energy storage devices (see, e.g., Ref. \cite{Zhao2012}).

\section{Acknowledgements}

This work was supported by the Flemish Science Foundation (FWO-Vl) and the Methusalem Foundation of the Flemish Government. M.N.-A was supported by the EU-Marie Curie IIF postdoc Fellowship/299855. One of us (F.M.P.) acknowledges discussions with Prof. Hongjun Gao. G.R.B acknowledges the support form the King Fahd University of Petroleum and Minerals, Saudi Arabia, under the TPRG131-CS-15 DSR project. A.C.T.vD acknowledges funding from AFOSR grants FA9550-10-1-0563 and  FA9550-11-1-0158.


\begin{thebibliography}{0}

\bibitem{Novoselov2005}K. S. Novoselov, A. K. Geim, S. V. Morozov, D. Jiang, M. I. Katsnelson, I. V. Grigorieva, S. V. Dubonos, and A. A. Firsov, Nature (London) {\bf 438}, 197 (2005).
\bibitem{Castro2009}A. H. Castro Neto, F. Guinea, N. M. R. Peres, K. S. Novoselov, and A. K. Geim, Rev. Mod. Phys. {\bf 81}, 109 (2009).
\bibitem{Yin1984}M.T. Yin and M.L. Cohen, Phys. Rev. B {\bf 29}, 6996 (1984).


\bibitem{Takeda1994}Takeda and K. Shiraishi, Phys. Rev. B {\bf 50}, 14916 (1994).
\bibitem{Fagan2000} S. B. Fagan, R. J. Baierle, R. J. Mota, A. J. R. da Silva, and A. Fazzio, Phys. Rev. B {\bf 61}, 9994 (2000).

\bibitem{Zhang2005}Y. Zhang, Y. W Tan, H. L. Stormer, and P. Kim, Nature (London) {\bf 438}, 201 (2005).
\bibitem{Guzman2007}G. Guzm\'{a}n-Verri and L. Lew Yan Voon, Phys. Rev. B {\bf 76}, 075131 (2007).
\bibitem{Cahangirov2009}S. Cahangirov, M. Topsakal, E. Akturk, H. \c{S}ahin, and S. Ciraci, Phys. Rev. Lett. {\bf 102}, 236804 (2009).
\bibitem{SahinPRB} H. \c{S}ahin, S. Cahangirov, M. Topsakal, E. Bekaroglu, E. Akturk, R. T. Senger, and S. Ciraci, Phys. Rev. B {\bf 80}, 155453 (2009).
\bibitem{Houssa2010}M. Houssa, G. Pourtois, V.V. Afanas'ev, and A. Stesmans, Appl. Phys. Lett. {\bf 97}, 112106 (2010).

\bibitem{LiuPRB2011}C.-C. Liu, H. Jiang, and Y. Yao, Phys. Rev. B {\bf 84}, 195430 (2011).
\bibitem{LiuPRL2011}C.-C. Liu, W. Feng, and Y. Yao, Phys. Rev. Lett. {\bf 107}, 076802 (2011).
\bibitem{Drummond2012} N. D. Drummond and V. Zolyomi, and V. I. Fal'ko, Phys. Rev. B {\bf 85}, 075423 (2012).
\bibitem{Ezawa2012}M. Ezawa, Phys. Rev. Lett. {\bf 109}, 055502 (2012).

\bibitem{Ag}C. L\'{e}andri, H. Oughaddou, B. Aufray, J.M. Gay, G. Le Lay, A. Ranguis, and Y. Garreau, Surf. Sci. {\bf 601}, 262 (2007); P. D. Padova, C. Quaresima, P. Perfetti, B. Olivieri, C. Leandri, B. Aufray, S. Vizzini, and G. L. Lay, Nano Lett. {\bf 8}, 271 (2008); B. Aufray, A. Kara, S. Vizzini, H. Oughaddou, C. Lean-dri, B. Ealet, and G. L. Lay, Appl. Phys. Lett. {\bf 96}, 183102 (2010); P. D. Padova, C. Quaresima, C. Ottaviani, P. M. Sheverdyaeva, P. Moras, C. Carbone, D. Topwal, B. Olivieri, A. Kara, H. Oughaddou, B. Aufray, and G. Le Lay, Appl. Phys. Lett. {\bf 96}, 261905 (2010); P. D. Padova, C. Quaresima, B. Olivieri, P. Perfetti, and G. L. Lay, Appl. Phys. Lett. {\bf 98}, 081909 (2011); C.L. Lin, R. Arafune, K. Kawahara, N. Tsukahara, E. Minamitani, Y. Kim, N. Takagi, and M. Kawai, Appl. Phys. Express {\bf 5}, 045802 (2012); B. Feng, Z. Ding, S. Meng, Y. Yao, X. He, P. Cheng, L. Chen, and K. Wu, Nano Lett. {\bf 12}, 3507 (2012); P. Vogt, P.D. Padova, C. Quaresima, J. Avila, E. Frantzeskakis, M.C. Asensio, A. Resta, B. Ealet, and G.L. Lay, Phys. Rev. Lett. {\bf 108}, 155501 (2012); L. Chen, C. Liu, B. Feng, X. He, P. Cheng, Z. Ding, S. Meng, Y. Yao, and K. Wu, Phys. Rev. Lett. {\bf 109}, 056804 (2012); H. Fnriquez, S. Vizzini, A. Kara, B. Lalmi, and H. Oughaddou, J. Phys.: Condens. Matter {\bf 24}, 314211 (2012); L. Chen, H. Li, B. Feng, Z. Ding, J. Qiu, P. Cheng, K. Wu, and S. Meng, Phys. Rev. Lett. {\bf 110}, 085504 (2013).
\bibitem{Zr}A. Fleurence, R. Friedlein, T. Ozaki, H. Kawai, Y. Wang, and Y. Yamada-Takamura, Phys. Rev. Lett. {\bf 108}, 245501 (2012).
\bibitem{Ir}L. Meng, Y. Wang, L. Zhang, S. Du, R. Wu, L. Li, Y. Zhang, G. Li, H. Zhou, W. A. Hofer, and H.-J. Gao, Nano Lett. {\bf 13}, 685 (2013).

\bibitem{Lin2013}C.-L. Lin, R. Arafune, K. Kawahara, M. Kanno, N. Tsukahara, E. Minamitani, Y. Kim, M. Kawai, and N. Takagi, Phys. Rev. Lett. {\bf 110}, 076801 (2013).
\bibitem{Guo2013} Z.-X. Guo, S. Furuya, J.-I. Iwata, and A. Oshiyama, Phys. Rev. B {\bf 87}, 235435 (2013).
\bibitem{Wang2013}Y.-P. Wang and H.-P. Cheng, Phys. Rev. B {\bf 87}, 245430 (2013).
\bibitem{Cahangirov2013}S. Cahangirov, M. Audiffred, P. Tang, A. Iacomino, W. Duan, G. Merino, and A. Rubio, Phys. Rev. B {\bf 88}, 035432 (2013).

\bibitem{Zhang2010}Y. Zhang and R. Tsu, Nanoscale Res. Lett. {\bf 5}, 805 (2010).

\bibitem{Kaloni} T. P. Kaloni, M. U. Kahaly, Y. C. Cheng, and U. Schwingenschlogl, J. Mater. Chem. {\bf 22}, 23340 (2012).


\bibitem{Porezag}D. Porezag, T. Frauenheim, T. Kohler, G. Seifert, and R. Kaschner, Phys. Rev. B {\bf 51}, 12947 (1995).

\bibitem{Elstner} M. Elstner, D. Porezag, G. Jungnickel, J. Elstner, M. Haugk, T. Frauenheim, S. Suhai, and G. Seifert, Phys. Rev. B {\bf 58}, 7260 (1998).

\bibitem{Frauenheim}T. Frauenheim, G. Seifert, M. Elstner, T. Niehaus, C. Kohler, M. Sternberg, Z. Hajnal, A. Di Carlo, and S. Suhai, J. Phys.: Condens. Matter {\bf 14}, 3015 (2002).

\bibitem{Gao_nature} J. Gao and J. Zhao, Scientific Reports {\bf 2}, 861 (2012).

\bibitem{Berger2006}C. Berger, Z. Song, X. Li, X. Wu, N. Brown, C. Naud, D. Mayou, T. Li, J. Hass, A. N. Marchenkov, E. H. Conrad, P. N. First, and
W. A. de Heer, Science {\bf 312}, 1191 (2006).

\bibitem{Singh2011}V. Singh, D. Joung, L. Zhai, S. Das, S. I. Khondaker, and S. Seal, Progress in Materials Science {\bf 56}, 1178 (2011).
\bibitem{Xia2012}C. Xia, S. Watcharinyanon, A. A. Zakharov, R. Yakimova, L. Hultman, L. I. Johansson, and C. Virojanadara, Phys. Rev. B {\bf 85}, 045418 (2012).
\bibitem{Wang2012} F. Wang, K. Shepperd, J. Hicks, M. S. Nevius, H. Tinkey, A. Tejeda, A. Taleb-Ibrahimi, F. Bertran, P. Le Fevre, D. B. Torrance, P. N. First, W. A. de Heer, A. A. Zakharov, and E. H. Conrad, Phys. Rev. B {\bf 85}, 165449 (2012).

\bibitem{Jose2011}D. Jose and A. Datta, Phys. Chem. Chem. Phys. {\bf 13}, 7304 (2011).
\bibitem{Sofo2007}J. O. Sofo, A. S. Chaudhari, and G. D. Barber, Phys. Rev. B {\bf 75}, 153401 (2007).

\bibitem{vanDuin} A. C. T. van Duin, S. Dasgupta, F. Lorant, and W. A. Goddard, J. Phys. Chem. A {\bf 105}, 9396 (2001).
\bibitem{Newsome}D. A. Newsome, D. Sengupta, H. Foroutan, M. F. Russo, and A. C. T. van Duin, J. Phys. Chem. C {\bf 116}, 16111 (2012).

\bibitem{PRB2013} S. K Singh, S. Goverapet Srinivasan, M. Neek-Amal, S. Costamagna, Adri C. T. van Duin, and F. M.
Peeters, Phys. Rev. B {\bf 87}, 104114 (2013).

\bibitem{Kresse} G. Kresse and J. Furthmuller, Phys. Rev. B {\bf 54}, 11169 (1996).
\bibitem{Zybin} S. V. Zybin, W. A. Goddard, P. Xu, A. C. T. van Duin, and A. P. Thompson, Appl. Phys. Lett. {\bf 96}, 081918 (2010).


\bibitem{Akturk}E. Akturk, C Ataca, and S. Ciraci, Appl.Phys. Lett. {\bf 96}, 123112 (2010).



\bibitem{Xian} L. Xian and M. Y. Chou, J. Phys. D {\bf 45}, 455309 (2012).

\bibitem{Singh2013} D. P. Kosimov, A. A. Dzhurakhalov, and F. M. Peeters, Phys. Rev. B {\bf 78}, 235433 (2008); Phys. Rev. B {\bf 81}, 195414 (2010); S. K. Singh, M. Neek-Amal, and F. M. Peeters, Phys. Rev. B {\bf 87}, 134103 (2013).

\bibitem{Liu1998} B. Liu, Z.-Y. Lu, B. Pan, C.-Z. Wang, K.-M. Ho, A. A. Shvartsburg, and M. F. Jarrold, J. Chem Phys. {\bf 109}, 9401 (1998).

\bibitem{Mehdi} M. Neek-Amal, A. Sadeghi, G. R. Berdiyorov, and F. M. Peeters, Appl. Phys. Lett. {\bf 103}, 261904 (2013).

\bibitem{suppl_1} Supplimental online video: structural transformation of a Si$_{18}$ cluster intercalating bilayer graphene (with 1920 carbon atoms) when the temperature is decreased from 2000 K to 0.1 K with rate 20 K/ps in molecular dynamics simulations.

\bibitem{suppl_2} Supplemental online video: snapshots of a Si$_{24}$ cluster intercalating bilayer graphene (with 1920 carbon atoms) during the DFTB optimization.

\bibitem{details} The damping constants for temperature and pressure were 100 fs and 2 ps, respectively, and the time step was 0.5 fs in all simulations.

\bibitem{Zhao2012} X. Zhao, C. M. Hayner, M. C. Kung, and H. H. Kung, Adv. Energy Mater. {\bf 1}, 1079 (2011).

\end{thebibliography}
\end{document}